\documentclass[a4paper]{VisionStyle}
\usepackage{epsfig}

\begin{document}

\title{XMM-Newton and Wide Field Optical Imaging to crack the mystery
of unidentified gamma-ray sources}

\author{P.A.\,Caraveo\inst{1} \and A.\,De Luca\inst{1,2} \and N.\,La Palombara\inst{1} \and R.\,Mignani\inst{3} 
\and E.\,Hatziminaoglou \inst{3}
 \and R.\,Hartman\inst{4} \and D.J.\,Thompson\inst{4} \and
 G.F.\,Bignami\inst{5} }

\institute{
  Istituto di Astrofisica Spaziale e Fisica Cosmica, Via Bassini 15, I-20133 Milano, Italy
\and
  Universit\`a di Milano Bicocca, Dipartimento di Fisica, Piazza della Scienza 3, I-20126 Milano, Italy
\and
  ESO, Karl Schwarzschild Str.  2, D-85748 Garching bei M\"unchen,  Germany
\and
  NASA GSFC, Greenbelt MD, USA
\and
  Agenzia Spaziale Italiana, Via di Villa Grazioli, 23 I-00198 ROMA,
  Italy }

\maketitle

\begin{abstract}

The limited  angular resolution  of gamma-ray telescopes  prevents the
straight  identification of the  majority of  the sources  detected so
far.   While  the  only  certified  galactic  gamma-ray  emitters  are
associated with pulsars, more than 90\% of the nearly 100 low latitude
sources detected by EGRET on  the GRO lack an identification. The best
identification strategy  devised over the  years relies on the  use of
X-ray  and optical data  to single  out possible  counterparts, rating
them  on  the basis  of  their  extreme  $F_{x}/F_{opt}$. \\  Here  we
describe our
multiwavelength programme based on  the EPIC mapping of selected EGRET
sources  complemented  by the  optical  coverage  of  the same  fields
obtained  through  the  Wide Field  Imager  (WFI)  of  the ESO  2.2  m
telescope. The field  of view of the WFI, comparable  to that of EPIC,
will allow to directly cross-correlate  X and optical data speeding up
significantly the selection  of interesting candidates worth follow-up
optical (and X-ray) studies.

\end{abstract}

\section{Introduction}

The  third  EGRET  catalogue  (\cite{pcaraveo_WB1:h99}) of  high
energy gamma-ray sources contains 271 entries  as opposed to the
25 listed in the COS-B  one.  The low latitude ($b\le  10^{\circ}$),
presumably galactic, sources went  from 22  to 80, while  the
high latitude  ($b\ge 10^{\circ}$), presumably  extragalactic, ones
jumped from  3 to  181,  with blazars accounting for roughly half
of them. \\ Surprisingly enough, EGRET has done very  little to
clarify the nature  of the low  latitude, mostly galactic,
gamma-ray emitters.   While their number increased fourfold,
successful  identifications remain at  a meagre  $\le$ 10\% level
and rely  only on  the  time  signature of  the  gamma-ray
photons,  which unambiguously  associates  these  sources  with
pulsars.  \\  In  the remaining  cases,  both  the   poorer
statistics  and  the  uncertain positions of the gamma-ray
sources have hampered ``blind'' periodicity searches, while the
size of the gamma-ray error boxes (on average 1 sq deg) prevents
a direct  optical identification. For these reasons, the search
for counterpart(s) has  been pursued through the X-ray coverage
of  the fields followed  by the  optical study  of each  X-ray
source. \\Figure 1 summarizes the ``zoom in" approach devised
during the 20 year long chase which led to the identification of
Geminga (\cite{pcaraveo_WB1:bc96}). The Einstein Observatory was
used to image the gamma-ray error box of Geminga, one of the
brightest gamma-ray source in the galactic plane, shown in the
upper right panel together with the slighly fainter Crab pulsar. All
but one of the four sources discovered by the Imaging
Proportional Counter were readily optically identified (\cite{pcaraveo_WB1:bcl83}).
The one lacking identification, 1E0630+178, was later
observed with the High Resolution Imager which yielded a few
sq. arcsec error box. This was the target of CFH, Palomar and the
ESO 3.6m telescopes, eventually yielding the m$_{V}$ 25.5 G" as the
candidate counterpart. This identification was later confirmed by
the discovery of the source proper motion (\cite{pcaraveo_WB1:bcm93}) 
which was found to improve the time solution of
the gamma-ray photon (\cite{pcaraveo_WB1:mhc96}), providing a
direct link between the bright gamma-ray source and its faint
optical counterpart. The link between the X and the gamma-ray
behaviour was provided by the discovery of the pulsation in
X-rays (\cite{pcaraveo_WB1:hh92}) followed by the confirmation in the
gamma-ray domain. The lack of radio emission prompted \cite*{pcaraveo_WB1:cbt96} 
to classify Geminga as a radio quiet neutron star, the
first of its kind, characterized by a gamma-ray  yield  1,000
times higher than the X-ray one, which , in turn, is 1,000 times
higher than the optical one. Indeed, the identification
procedure  relies on  the extreme $F_{x}/F_{opt}$ value measured
for  Isolated Neutron  Stars (INSs),  6 of which are known  to be
gamma-ray emitters (see e.g. \cite{pcaraveo_WB1:t01}).  Such  a
multiwavelength  approach is now the standard  method to  search
for  the counterparts  of EGRET gamma-ray sources.

\begin{figure*}[t]
  \begin{center}
\epsfig{file=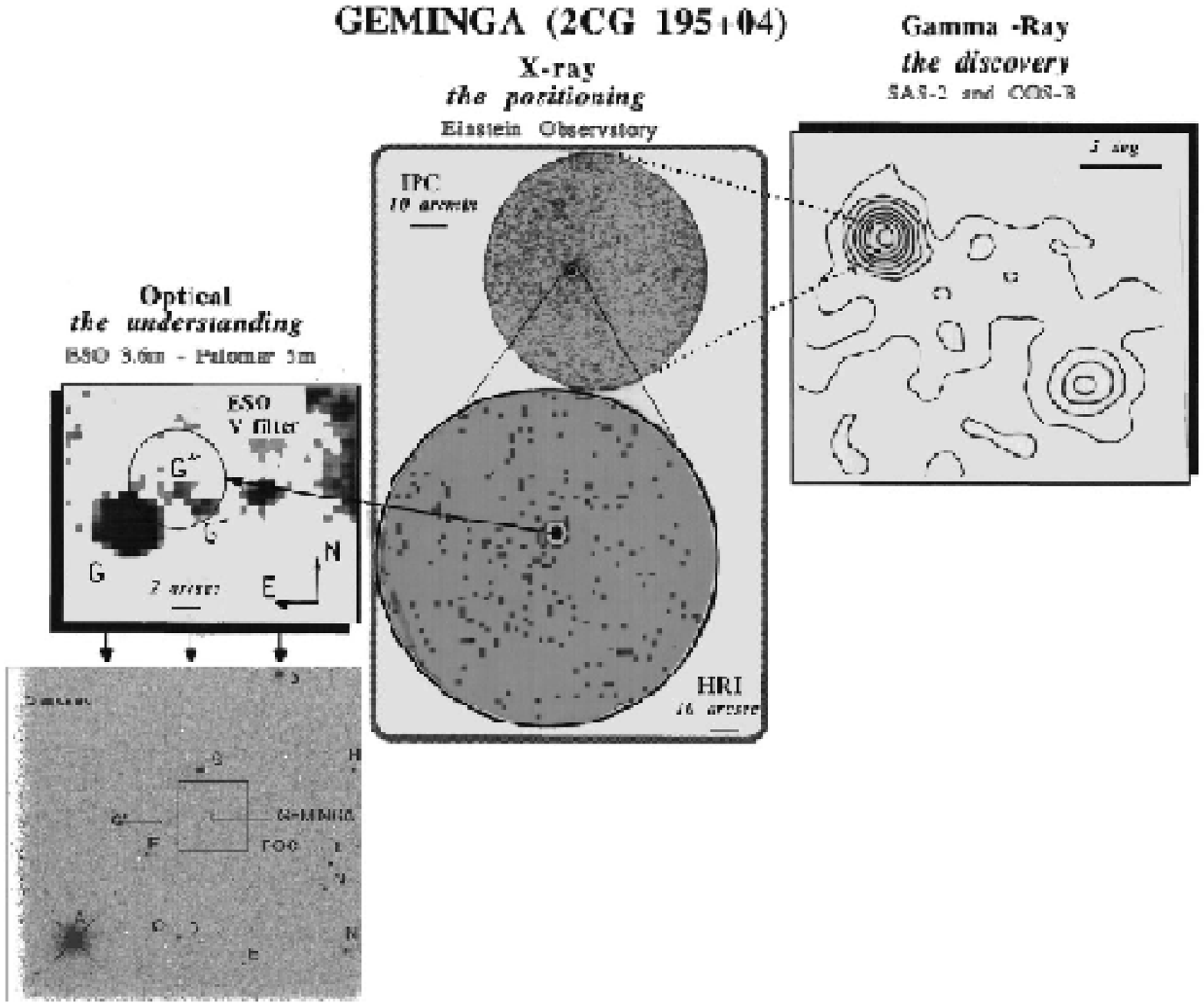, width=17cm}
  \end{center}
\caption{From upper right  to lower left: sketch of  the 20
years-long identification chain  which led to the identification
of Geminga. The gamma ray source, discovered by SAS-2 and
positioned by COS-B(upper right), was investigated using both the
low and high resolution instruments on board the Einstein
Observatory (center panel). The few arcsec error radius of the
most promising source was imaged at length by CFH, 5m Hale and
ESO telescope yielding the candidate optical counterpart
G''(upper left panel). G'' was then studied with the HST (lower
left panel).}
\end{figure*}

Years of  multiwavelength efforts  have yielded (so  far) less  than a
dozen tentative  identifications encompassing 4  energetic young radio
pulsars, 3 Geminga-like radio quiet isolated neutron stars, 2 peculiar
binary systems,  1 low  latitude Blazar and  few pulsars  nebulae (see
\cite{pcaraveo_WB1:c01} for a review of the proposed identifications).\\
Attempts to associate the low latitude unidentified EGRET sources with
different       classes      of       galactic       objects      (see
e.g. \cite{pcaraveo_WB1:rbt99}; \cite{pcaraveo_WB1:g00}) have not yielded,
so far,  gamma-ray source templates more appealing  than the classical
Isolated  Neutron   Star  one.   Thus,  while   not  neglecting  other
possibilities, it is natural to assume that at least a fraction of the
remaining low latitude sources are unidentified neutron stars. \\ Our
 EPIC-ESO programme, centered on two middle latitude  EGRET
sources, is precisely aiming at these objects.

\section{INSs: radio-loud vs. radio-quiet}

The search  for neutron stars in  gamma-rays started as  soon as COS-B
discovered    the    UGOs    (Unidentified    Gamma    Objects,    see
\cite{pcaraveo_WB1:bh83} for  a review), but  no new pulsar,  after Crab
and Vela, was  unveiled.  New searches have been  spurred by the EGRET
detections of four more  pulsars (e.g.  \cite{pcaraveo_WB1:t96}) but the
lack  of results,  experienced at  the time  of COS-B,  appears  to be
substantially      unchanged.       Dedicated      radio      searches
(\cite{pcaraveo_WB1:ns97}),  aimed  precisely at  the  search for  radio
pulsars inside the  error boxes of 10 of  the brightest EGRET sources,
yielded null  results, showing  that the straightforward  radio pulsar
identification is not the only  possible solution to the enigma of the
unidentified  high-energy gamma-ray  sources.  This  has  been further
strengthened  by the work  of \cite*{pcaraveo_WB1:n96}  who investigated
350  known  pulsars,  finding   few  positional  coincidences  but  no
significant gamma-ray timing  signature for any of the  pulsars in the
survey.   The   recently  released  portion  of   the  Parkes  survey,
encompassing 368 new  pulsars, has been correlated to  the EGRET error
boxes, finding  no more than  3 plausible pulsar candidates  against a
dozen  of  chance  coincidences  (\cite{pcaraveo_WB1:tbc01}).   With  no
gamma-ray  instruments in  operation,  it is  presently impossible  to
confirm these associations.  \\ Indeed, gamma-ray astronomy does offer
a remarkable  example of an  INS which behaves  as a pulsar as  far as
X-and-gamma-astronomy are  concerned but has little, if  at all, radio
emission.  As an established representative of the non-radio-loud INSs
(see \cite{pcaraveo_WB1:cbt96} for a  review), Geminga offers an elusive
template  behaviour:  prominent  in  high  energy  gamma-rays,  easily
detectable in X-rays and downright  faint in optical, with sporadic or
no radio  emission.  Although the energetics of  a Geminga-like object
is not adequate to account  for the very low latitude (presumably more
distant)  EGRET sources,  the third  EGRET catalogue  contains several
middle  latitude  sources  which  could  belong to  a  local  galactic
population.   In  this  case,   their  gamma-ray  yield  is  certainly
compatible with  the rotational energy  loss of a middle  aged neutron
star, like Geminga.

Thus, it makes  sense to apply to these  sources the two-step strategy
devised    during   the    20    year   long    chase   for    Geminga
(\cite{pcaraveo_WB1:bc96}), more  recently also applied  to pinpoint the
candidate neutron star counterpart  to the EGRET source 3EG J1835+5918
(\cite{pcaraveo_WB1:mh01}) as  well as to  few other EGRET  sources (see
\cite{pcaraveo_WB1:c01}). First of all, one  has to start from a list of
unidentified  X-ray sources detected  in the  EGRET error  boxes. Next
step  is  to  single  out  potential neutron  star  candidates  taking
advantage of their high $F_{x}/F_{opt}$ as a distinctive character and
using multicolor  information as a  further handle to  solve ambiguous
cases, e.g., when  more optical entries are compatible  with the X-ray
position.

\section{Our XMM-ESO programme}

\subsection{X-ray side}

Currently, we are focussing  on two middle-latitude EGRET sources
(3EG 0616-3310  and  3EG  1249-8330),   selected  on  the  basis
of  their relatively good positional accuracy, spectral shape,
galactic location and  lack of  candidate extragalactic
counterpart. EPIC (\cite{pcaraveo_WB1:tu01}; 
\cite{pcaraveo_WB1:s01}) is the ideal
instrument to perform the X-ray coverage of  the gamma-ray
error boxes, since it offers good
angular resolution ($\le$ 6" FWHM), coupled to high sensitivity in a broad 
energy range (0.1$\div$12 keV) and good spectral resolution ($E / \Delta E$ $\approx$ 
20$\div$50), over a wide field of view ($\approx$ 15 arcmin radius). 
Thus, each gamma-ray
source error box, a circle of 30 arcmin radius, can be covered
with four EPIC exposures. For a net exposure time of $10\,000$ s 
for each pointing, this yields an
homogeneous coverage of about 1 square degree  down to a
$5\sigma$  detection limiting  flux ranging from
$\approx7\times10^{-15}$        erg cm$^{-2}$~s$^{-1}$ to
$\approx2\times10^{-14}$  erg cm$^{-2}$~s$^{-1}$,  depending on
the source spectrum.   Taking these values  as reference, folded
with the $Log    N-Log   S$ function    for   extragalactic
sources   (see e.g. \cite{pcaraveo_WB1:h01})  and   including  an
evaluation  of  the galactic  contribution  (see  e.g.
\cite{pcaraveo_WB1:g96}),  we were expecting $\simeq 100$  sources
(mainly AGN and active stars), positioned to within 5 arcsec, in
each EGRET error box. To quickly identify such a number of
serendipitous sources, we had to devise an ad hoc optical
strategy.

\subsection{The optical side}

Given the range of $F_{x}/F_{opt}$ values characteristic for the
known classes of X-ray sources  (see, e.g.,
\cite{pcaraveo_WB1:k99}), we ought to reach $V \simeq 25$ in the
optical follow-up in order to be able to discard  most non-INSs
identifications. Thus,  although useful  for a first filtering,
the available Digital Sky Surveys are not deep enough for our
purpose: dedicated optical observations are needed.\\
However, the expected EPIC yield prompted us to seek an approach
different from the one-by-one philosophy used so far to identify
the few sources  detected by previous X-ray telescopes in each
gamma-ray error box.
 It is apparent that without a``massive" approach to the identification work the optical side is
bound  to became the bottleneck  of our a multiwavelength chain.
Our optical work should be aimed at dozens of faint objects
distributed over about 1 sq deg. A dedicated, deep, multicolour
optical \underline{survey} of our fields must be the starting
point for the real identification work. This  is why we have
decided to rely on the WFI of the 2.2 m as the optical complement
of our EPIC coverage of EGRET sources. The WFI large field of
view ($33 \times 34$ arcmin), directly comparable to the  EPIC
one, will allow to cross correlate directly X  and optical  data
\underline{speeding up dramatically the identification work}.

\subsection{The state of the art}

The XMM/EPIC  observation campaign  of the two  EGRET error
boxes was completed in November  2001. The optical data are  now
being collected at the ESO 2.2m telescope (Period 68).  The
analysis of the X-ray data is currently in progress, using the
XMM-Newton Science Analysis System (XMM-SAS) v5.2. \\
Unfortunately, we discovered  that  space  weather  was  not
favourable during  2  of  our  8 pointings.   These observations
(1 per EGRET error  box)  were badly affected  by high  particle
background episodes  (the so-called  soft proton flares, see XMM
Users' Handbook) and we were forced to reject up to 80\% of the integration
time. \\ We  ran a preliminary source  detection over the broad
energy  band 0.3-8 keV, which gathers  almost all the usable
X-ray events from celestial sources. About 130 sources per EGRET
error box were found and their  fluxes, computed using a simple
Geminga-like spectrum  (a   black  body  with  kT=0.1   keV  and
a   Nh  of  order 5$\times$10$^{20}$ cm$^{-2}$), range  from
$\approx 3 \times 10^{-13}$ erg cm$^{-2}$ s$^{-1}$ to $\approx$
10$^{-14}$ erg cm$^{-2}$ s$^{-1}$.
\\ Figure 2  shows the  EPIC mosaic  of the field  of 3EG  1249-8330.
The detected sources are marked  with little circles. The South
West observation clearly shows the dramatic effect of flaring
particle background.   A more sophisticated source detection
analysis is currently  under way.  Following \cite*{pcaraveo_WB1:m01}, 
particular care  is now
devoted  to create a background map which properly reproduce the
background fluctuations on small spatial scales,  a crucial step
in  order to  assess the reality and the significance of our
sources.

\begin{figure*}[t]
  \begin{center}
\epsfig{file=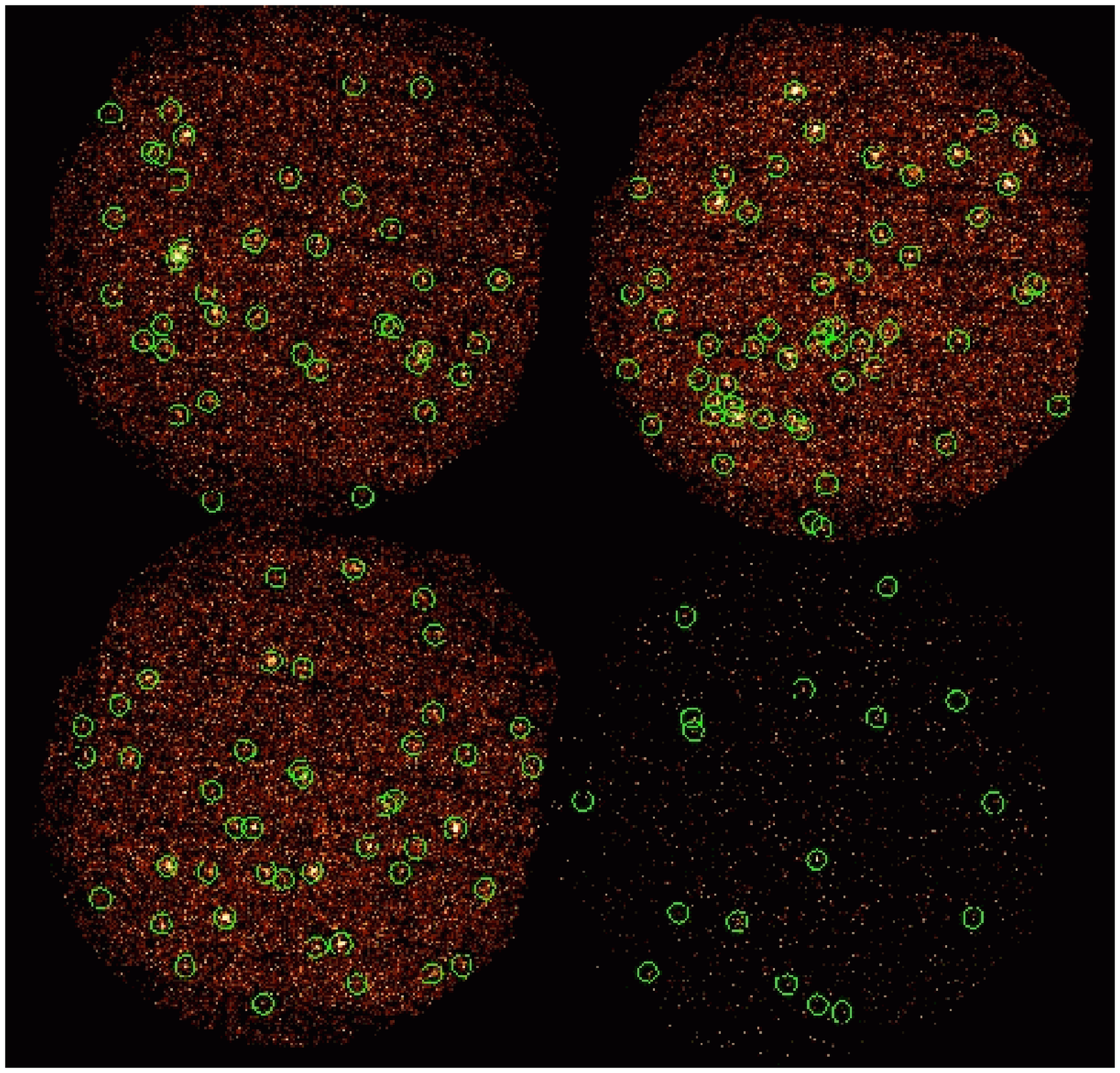, width=17cm}
  \end{center}
\caption{EPIC mapping of the error  box of 3EG 1249-8330. North is up,
East is left.   The detected sources, marked with  circles, are more
than 130.   The lower  right pointing is  the one affected  by flaring
background. The reduction of the fraction of good observing time is evident.}
\end{figure*}

 As
a first screening to filter out obvious non-INS identifications,
we take advantage  of the available optical  catalogues  to cross
correlate our lists of sources (the position of which is known
within 5 arcsec).  Our starting point is the extended
version of the recently released Guide Star Catalogue 2 (GSC2),
an all-sky, multi-epoch and multicolor optical catalogue based
on photographic surveys carried out between 1953 and 1991 (McLean
et al.  2002-in preparation).  The GSC2 provides color
information in at least three photographic passbands  (roughly
corresponding to $B$, $R$ and $I$)  down to $B \sim 22$  and
morphological classification at a $\sim 90  \%$ confidence level
for  objects in our  latitude range and brighter  than $B  \sim
19$.  To extend  our color coverage to  the infrared ($J$, $H$
and $K$) we use the updated release of the  2 Micron All Sky
Survey (2MASS) catalogue.
\\ After the  cross  correlations, the color catalogues   are
combined and  used   as  input   for the object classification
procedure, with the morphological classification from the GSC2 as
a further aid.  For this, we apply the automatic algorithm
developed  by \cite{pcaraveo_WB1:h00}, which   fits model fluxes,
simulated  from template  spectral libraries, to the observed
ones through  a $\chi^2$  minimization technique. The reliability
of this method has been  tested successfully by
\cite{pcaraveo_WB1:h02} for the classification of  the objects
detected in the  Chandra Deep Field South.  Finally,  the
properties of  the X-ray sources  (e.g. hardness ratios,
spectra)  will be compared  with the classifications  of their
candidate optical counterparts and identifications evaluated
through a decision tree.
\\ The  whole procedure  will be  then  repeated for  the remaining,  non
identified,  EPIC sources, using  deeper optical  catalogues extracted
from the WFI data. The observations (in $UBVRI$) are now being carried
out in Service Mode  by the 2.2m team.  So far, only  the field of 3EG
0616-3310  has been  observed, with  an  area coverage  of $\simeq  25
\%$. \\  Of course,  only the availability  of the  multicolor optical
data  from  WFI  will allow  us  to  perform  the  final step  of  our
identification chain.

\section{Conclusions }
The identification of gamma-ray sources, both individually
and as a population, rests on multiwavelength observations.
The programme we are pursuing at X-ray and optical wavelengths
will yield, in a relatively short time, candidate counterparts
worth follow-up investigations.
Moreover, we plan to make available the complete catalog of
the sources detected in our 1.6 sq. deg medium
galactic latitude survey.

\begin{acknowledgements}

This work is supported by the Italian Space Agency (ASI).

\end{acknowledgements}

\end{document}